\documentclass[aps,onecolumn]{revtex4}

\usepackage{bm}
\usepackage[]{graphicx}
\usepackage{amsmath}
\usepackage{amssymb}
\usepackage{color}

\begin{document}

\title{Measuring Oscillatory Velocity Fields Due to Swimming Algae}

\author{Jeffrey S. Guasto$^{1}$, Karl A. Johnson$^{2}$, and J.P. Gollub$^{1,3}$}
\affiliation{$^{1}$Department of Physics, Haverford College, Haverford, Pennsylvania 19041, USA}
\affiliation{$^{2}$Department of Biology, Haverford College, Haverford, Pennsylvania 19041, USA}
\affiliation{$^{3}$Department of Physics, University of Pennsylvania, Philadelphia, Pennsylvania 19104, USA}

\date{\today}

\begin{abstract}
In this fluid dynamics video, we present the first time-resolved measurements of the oscillatory velocity field induced by swimming unicellular microorganisms.
Confinement of the green alga \emph{C. reinhardtii} in stabilized thin liquid films allows simultaneous tracking of cells and tracer particles. 
The measured velocity field  reveals complex time-dependent flow structures, and scales inversely with distance. 
The instantaneous mechanical power generated by the cells is measured from the velocity fields and peaks at 15 fW.  
The dissipation per cycle is more than four times what steady swimming would require.  
\end{abstract}

\maketitle

\section*{Submission Contents}

There are two videos contained in this submission for the APS DFD 2010 Meeting Gallery of Fluid Motion in Long Beach, CA both in MPEG-4 format:
\begin{itemize}
	\item \textbf{Guasto\_APSDFD2010\_Video1.mp4} - high quality video (26.2 MB) appropriate for the DFD 2010 Gallery of Fluid Motion display.
	\item \textbf{Guasto\_APSDFD2010\_Video2.mp4} - low quality video (8.0 MB) appropriate for web viewing.
\end{itemize}

\section*{Acknowledgement}

We thank R.E. Goldstein, K. Drescher, E. Lauga, and M.D. Graham for helpful discussions, as well as B. Boyes for technical assistance. This work was supported by NSF Grant DMR-0803153.


\end{document}